\documentclass[prc,showpacs,floatfix]{revtex4}
\usepackage{bm}
\usepackage{graphicx}
\usepackage{amsmath}
\usepackage{bbold}

\newcommand{\beq}{\begin{equation}}
\newcommand{\eeq}{\end{equation}}
\newcommand{\beqa}{\begin{eqnarray}}
\newcommand{\eeqa}{\end{eqnarray}}

\begin{document}

\title{Transverse electron scattering response function of $^3$He in the quasi-elastic peak 
region and beyond with $\Delta$ isobar degrees of freedom
}

\author{Luping Yuan$^{1}$,
  Winfried Leidemann$^{1,2}$,
  Victor D. Efros$^{3}$,
  Giuseppina Orlandini$^{1,2}$,
  and Edward L. Tomusiak$^{4}$
        }

\affiliation{
 $^{1}$Dipartimento di Fisica, Universit\`a di Trento, I-38123 Trento, Italy \\
 $^{2}$Istituto Nazionale di Fisica Nucleare, Gruppo Collegato di Trento,
 I-38123 Trento, Italy \\
 $^{3}$European Centre for Theoretical Studies in Nuclear Physics and Related Areas(ECT*),
 Villa Tambosi, I-38123 Villazzano (Trento), Italy\footnote{On leave from National Research Centre
 "Kurchatov Institute",  123182 Moscow,  Russia}\\
 $^{4}$Department of Physics and Astronomy,
 University of Victoria, Victoria, BC V8P 1A1, Canada\\
}

\date{\today}
\begin{abstract}
The $^3$He transverse electron scattering response function $R_T(q,\omega)$ is calculated in the quasi-elastic peak region and beyond for momentum transfers $q$ = 500, 600 and 700 MeV/c.  Distinct from our previous work for these kinematics where we included meson exchange currents and relativistic corrections we now additionally include $\Delta$ isobar currents ($\Delta$-IC). The $\Delta$-IC contribution increases the quasi-elastic peak height by about 5\% and leads to an excellent agreement with experimental data in the whole peak region. In addition it is shown that effects due to the three-nucleon force largely cancel those due to the $\Delta$-IC in the peak region. Finally, we have found that $\Delta$-IC are important for three-body break-up reactions in the so-called dip region. This could explain why in a previous study of such a reaction, where $\Delta$ degrees of freedom were not included, no agreement between experimental and theoretical results could be obtained.

\end{abstract}

\maketitle

It is well known that $\Delta$ degrees of freedom play an important role in the response of the two-nucleon system to virtual photons (see e.g. \cite{ALT05}). For the three-nucleon system a study of $\Delta$ effects in inclusive electron scattering was made in \cite{Deltuva04}. Large effects were found at higher momentum transfer close to the break-up threshold. In \cite{Deltuva04} the quasi-elastic peak region was also studied at $q \le 500$ MeV/c, but $\Delta$ degrees of freedom had  only a marginal influence.
In particular at $q=500$ MeV/c almost no $\Delta$ effect was found. In the present work we study the effect of $\Delta$-IC on the transverse response function $R_T(q,\omega)$ in the quasi-elastic region for somewhat higher momentum transfers, i.e. 500 MeV/c $\le q \le$ 700 MeV/c. Our calculation is performed with full consideration of the final state interaction by applying the Lorentz integral transform method \cite{LIT}. In previous studies for this kinematics we have shown that relativistic effects are important, whereas meson exchange current contributions are  small \cite{ELOT10,ELOT11}. Our previous results for $R_T$ are in close agreement with experimental data although they slightly underestimate the experimental quasi-elastic peak height. With the present inclusion of isobar currents we further improve the description of the nuclear current operator. This enables us to check whether an even better agreement with experiment can be obtained with the additional $\Delta$-IC.

The $\Delta$-IC are calculated in impulse approximation (IA) as described in \cite{YELT10}. Here we only give a short summary. We split the $^3$He ground-state wave function $\Psi_0$ and the Lorentz state $\tilde\Psi$ into  NNN and NN$\Delta$ parts, i.e.
\begin{equation}
|\Psi_0\rangle = |\Psi_0^N\rangle  + |\Psi_0^{\Delta}\rangle \,,  \,\,\,\,\,\,\,\,\,
|\tilde\Psi\rangle = |\tilde\Psi^N\rangle  + |\tilde\Psi^{\Delta}\rangle \,.
\end{equation}
Then $\Psi_0^N$ is calculated by solving the Schr\"odinger equation with a Hamiltonian $H_N$ which contains a realistic nuclear potential consisting of a two- and a three-nucleon force. In a next step $\Psi_0^{\Delta}$ is determined in IA by using the calculated $\Psi_0^N$. Finally, the following LIT equation is solved
\begin{eqnarray}
\label{lorentzstate1} ( H_N  - E_0  - \sigma ) |\tilde\Psi ^N  \rangle &=& - V^{NN,N\Delta}
 ( H_{\Delta}  - E_0 - \sigma)^{-1} \left( {O}_{\Delta N}  |\Psi ^N _0 \rangle  + {O}_{\Delta\Delta}
  |\Psi ^{\Delta} _0 \rangle \right)  \nonumber \\
&& + {O}_{NN}  |\Psi ^N _0 \rangle  + {O}_{N\Delta}  |\Psi ^{\Delta} _0 \rangle \label{newschrlorn} \,,
\end{eqnarray} 
where $E_0$ is the three-body ground-state energy, the complex $\sigma=\sigma_R + i \sigma_I$
is the argument of the LIT in the transformed space, the ${O}_{N_1 N_2}$ denote the
various diagonal (N$_1$=N$_2$) and transition (N$_1\ne $N$_2$) electromagnetic current operators,
$V^{NN,N\Delta}$ is the transition potential from NN to N$\Delta$, and $H_{\Delta}$
denotes the diagonal Hamiltonian of the NN$\Delta$ channel, where we include the $N$-$\Delta$ mass difference, $\delta m$ = $M_\Delta-M_N$, and the kinetic energy. The norm of the Lorentz state $\tilde\Psi$ leads to the LIT of $R_T$. The response function can then be obtained by inversion of the transform.

Our previous studies of the transverse quasi-elastic response used a nuclear current operator \cite{ELOT10}  which  only included the non-relativistic nucleon one-body current with first-order relativistic corrections and a two-body current (MEC). Moreover,
we have tried to minimize additional but not explicitly treated relativistic effects by performing the calculation in a specific reference frame, namely the active nucleon Breit (ANB) frame \cite{ELOT05}. In this frame all nucleons in the target move with -${\bf q}/2$, i.e. the target nucleus has a initial momentum ${\bf p}_i = -A{\bf q}/2$. As opposed to non-relativistic calculations in all other frames, an ANB frame calculation, with results properly transformed to the laboratory (lab) frame, leads to the correct description of the experimental quasi-elastic peak position. In addition, both for longitudinal \cite{ELOT05} and transverse responses \cite{ELOT11}, we have shown that the rather large frame dependence can be significantly reduced by introducing a quasi-elastic two-fragment break-up model which allows the use of proper relativistic two-body kinematics while having no effect on the dynamical calculation. Applying this model to the ANB frame gives no effect on the peak position, whereas the peak height is slightly increased. For other frames the two-body break-up model shifts the peak position to the correct position in the lab frame but still leaves some frame dependence in the quasi-elastic peak height. As pointed out in \cite{ELOT11} the ANB frame result should be the most reliable one, since there are profound reasons to expect the smallest relativistic corrections in this frame. This is confirmed by the calculated size of the explicitly treated relativistic corrections for the various frames \cite{ELOT11}. 

In the present study we use the Argonne V18 NN potential \cite{AV18} and the Urbana IX three-nucleon force (3NF) \cite{UIX}. All calculations are made in the ANB frame and the resulting  $R^{\rm ANB}_T(q_{\rm ANB},\omega_{\rm ANB})$ is transformed to obtain the laboratory (lab) frame result $R_T(q,\omega)$. Further details of the calculations are described in \cite{ELOT10}. 
In contrast to \cite{ELOT10,ELOT11} we include here the above described $\Delta$-IC. This contribution is taken into account for all transitions to final states of the three-nucleon system with total angular momentum $J_f \le 15/2$, whereas the nucleon one-body current operator is evaluated up to higher $J_f$ as described in \cite{ELOT10} fulfilling quite well the non-relativistic sum rule \cite{ELOT11}. 

\begin{figure}[ht]
\centerline{\includegraphics*[scale=.5]{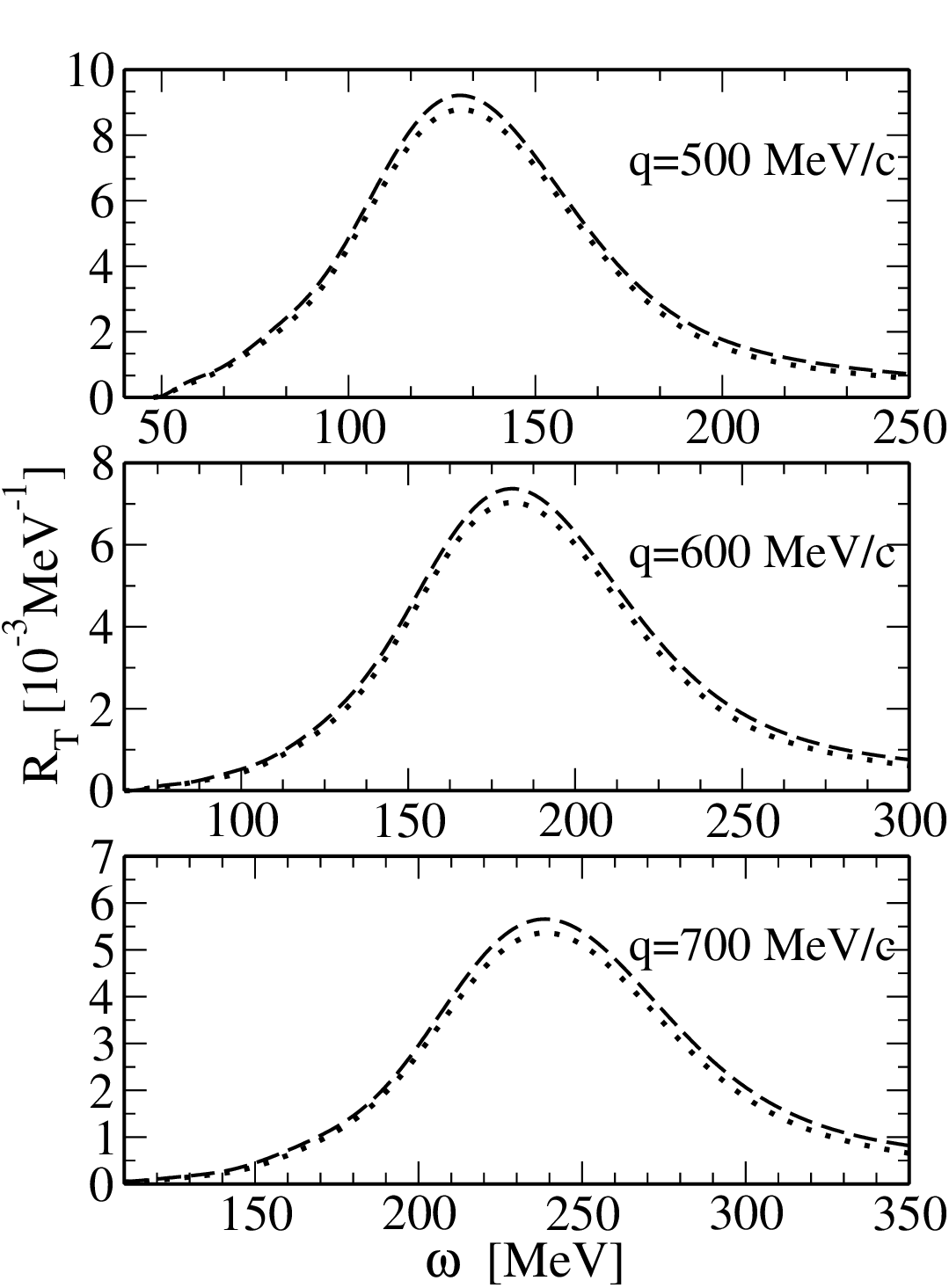}}
\caption{$R_T(q,\omega)$ without (dotted) and with (dashed) $\Delta$-IC contribution.}
\label{IC}
\end{figure}
In Fig.~1 we show two results for $R_T(q,\omega)$, one with and one without the $\Delta$-IC contribution, while the non-relativistic one-nucleon current with first-order relativistic corrections and a meson exchange current are included in both cases. One sees that the $\Delta$ isobar currents lead to an overall increase of $R_T$.  The quasi-elastic peak height is moderately enhanced by about 5\%, whereas relative increases are somewhat larger at higher energies.

Our calculation can be further improved by using the above mentioned kinematical two-fragment model. The result is illustrated in Fig.~2 
where one notes a slight increase of the peak height on top of that already produced by the $\Delta$-IC contribution. It is evident that  inclusion of the $\Delta$-IC contribution provides excellent agreement with experimental data in the quasi-elastic peak region for all three momentum transfers.
Here we should not forget to mention that also the relativistic corrections to the one-body current operator, which were not considered in \cite{Deltuva04}, gives a not unimportant contribution to $R_T$ at $q \ge 500$ MeV/c.
For a detailed discussion of this contribution we refer to \cite{ELOT10, ELOT11}.
			       
\begin{figure}[ht]
\centerline{\includegraphics*[scale=.5]{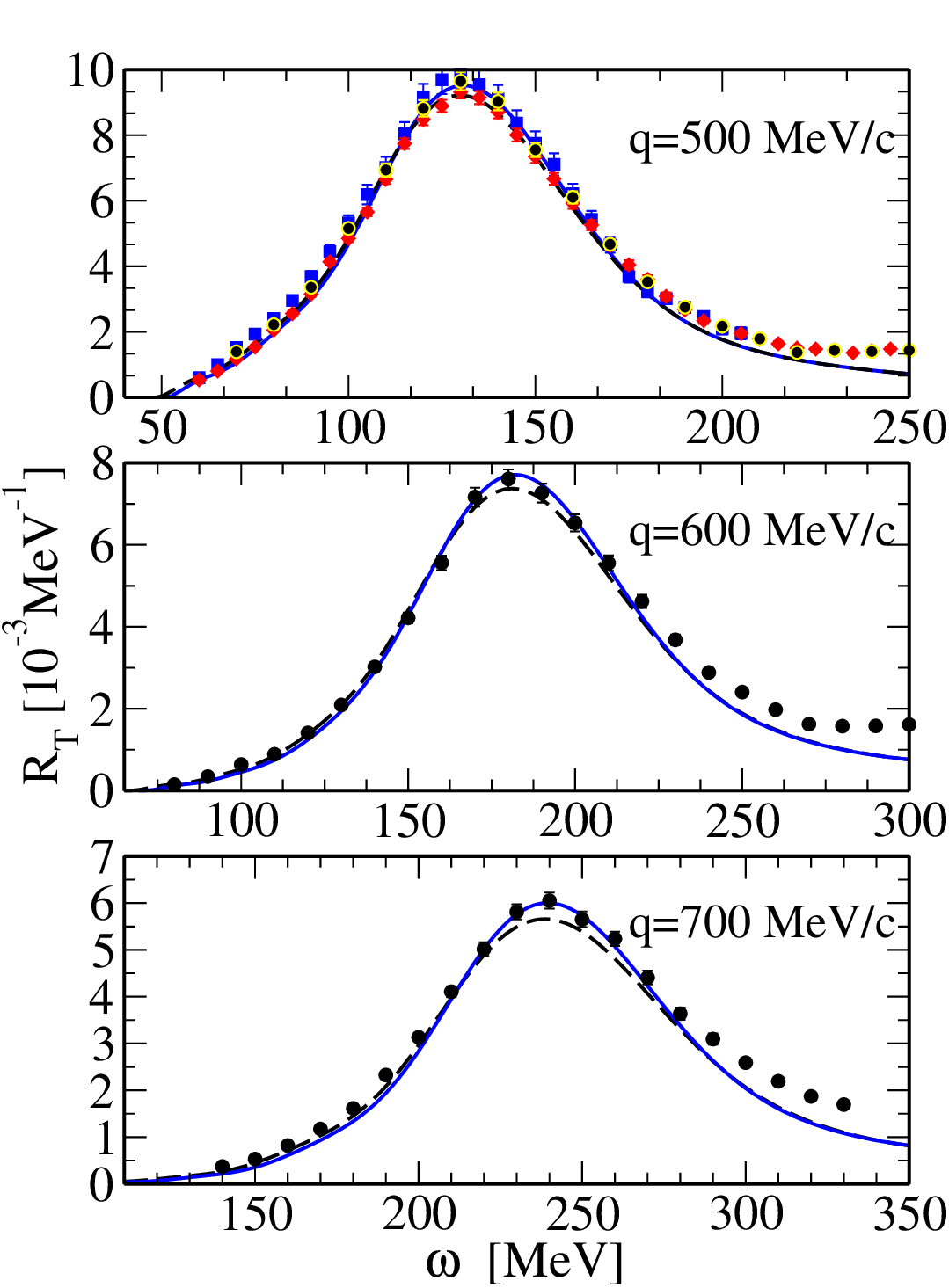}}
\caption{$R_T(q,\omega)$: dashed curve as in Fig.~1, full curve represents
the result of a calculation with the same theoretical ingredients but using the kinematical two-fragment model (see text). Experimental data from \cite{Saclay} (squares), \cite{Bates} (diamonds), \cite{world} (circles).}
\label{exp}
\end{figure}

As mentioned in the introduction almost no $\Delta$ effects were found by \cite{Deltuva04} in the quasi-elastic peak region at $q=500$ MeV/c. This appears to contradict our result shown in Fig.~1 for $R_T$ at $q=500$ MeV/c. Before we come to a clarification of this we should mention that the calculation of \cite{Deltuva04} is a coupled channel calculation with N and $\Delta$ degrees of freedom, which, below pion threshold is in principle a more consistent treatment than our IA approach.
For such a coupled channel calculation it is correct not to take into account a 3NF resulting from $\Delta$ degrees of freedom, since the $\Delta$ channel affects the nucleonic channel
via a transition potential. This differs from the IA where an explicit consideration of a 3NF is necessary.
Above pion threshold explicit pion degrees of freedom are missing in both calculations. 
However, they both should represent rather reasonable approximations even above pion threshold as long as the internal energy transfer to the three-nucleon system, $\omega_{\rm int}$, remains sufficiently below the $N$-$\Delta$ mass difference $\delta m$.

From the discussion above it is evident that in order to compare $\Delta$ contributions in our results with those of \cite{Deltuva04} we have to combine the 3NF and $\Delta$-IC contributions of our calculation. 
In Fig.~3 we show the separate effects on $R_T^{\rm ANB}$ due to the hadronic (3NF effect) and the electromagnetic ($\Delta$-IC effect) interaction (we choose the ANB frame for this comparison, since it is more convenient for us). One sees that the 3NF reduces the quasi-elastic peak height by about 5\%, whereas, as mentioned before, $\Delta$-IC lead to an increase by the same percentage. In fact for $q_{\rm ANB}=$500 and 600 MeV/c one finds a nearly perfect cancellation of both effects in the whole peak region, while at $q_{\rm ANB}$=700 MeV/c the $\Delta$-IC contribution is a bit larger than the 3NF effect. At higher energies both effects increase $R_T$. On the other hand, as pointed out above, our calculation becomes less realistic beyond pion threshold and our treatment should become increasingly inadequate with further growing energy. Nonetheless we think that our calculation leads at least to a reasonable estimate of $R_T$ up to about $\omega_{\rm int}$=250 MeV. For the two higher $q_{\rm ANB}$-values we show results at even higher energies, but one should be aware that there our calculation has only a rather qualitative value. However, even the energies displayed are not yet in the regime of quasi-elastic $\Delta$ knockout, which is located in the lab frame near $\omega = \delta M + q^2/2M_\Delta$ leading to the following values for $\omega_{\rm int}$: 350, 375, and 405 MeV for $q=500$, 600, and 700 MeV/c, respectively (note that in the quasi-elastic peak region $q_{\rm ANB}$ is somewhat smaller than the corresponding properly Lorentz transformed $q$ value). 
\begin{figure}[ht]
\centerline{\includegraphics*[scale=.5]{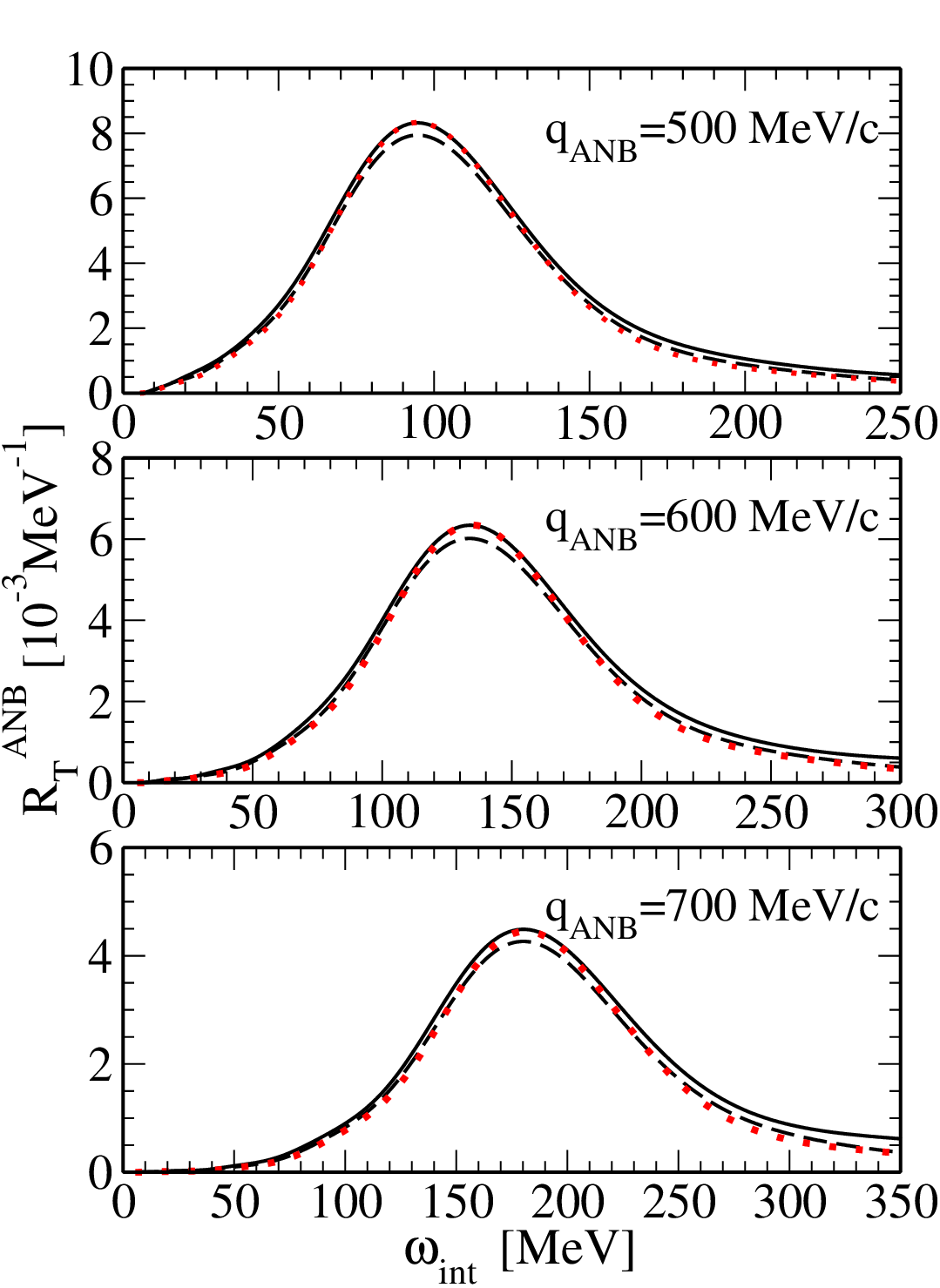}}
\caption{$R_T^{\rm ANB}$ as function of internal excitation energy $\omega_{\rm int}$ of the three-nucleon system: no $\Delta$-IC and no 3NF (dotted), no $\Delta$-IC but 3NF included (dashed), both $\Delta$-IC and 3NF included (full). MEC are not taken into account for any of the three curves.}
\label{T_all}
\end{figure}

In Fig.~4 we make the same comparisons as in Fig.~3, but only for transitions to final states with a total isospin of $T_f=3/2$. The figure shows that for this isospin channel there is a large $\Delta$ effect in the so-called dip region.  This arises mainly from the $\Delta$-IC and to a lesser extent from the 3NF. The total  effect amounts to the following increases of $R_T$ at $\omega_{\rm int} = 250$ MeV (in parentheses the results for the $T_f=1/2$ channel): 82\% (33\%), 45\% (26\%), and 21\% (13\%) at $q_{\rm ANB}=500$, 600, and 700 MeV/c, respectively. This finding is very interesting, since the $T_f=3/2$ channel contributes exclusively to the three-body break-up. Because of the considerably lower increases for the $T_f=1/2$ channel, where both two- and three-body break-up are possible, one could speculate that also for this channel mainly the three-body break-up reaction is affected.  We conclude that $\Delta$ degrees of freedom should be of greater importance for the $^3$He$(e,e'pp)$ and $^3$He$(e,e'pn)$ reactions in the dip region. Here it is worthwhile mentioning that  recently the reaction $^3$He$(e,e'pn)$ has been studied in the dip region for various momentum transfers $q$ ranging from 300 to 450 MeV/c \cite{Middleton09}. Rather large differences were found between experimental and theoretical results, but neither a 3NF nor $\Delta$-IC were included in the theoretical calculation. Though the momentum transfers are a bit lower than in our study one can infer from the $q$-dependence of
our results that the large $\Delta$ effect will be quite significant also for lower $q$. Therefore a consideration of $\Delta$ degrees of freedom in a calculation of $^3$He$(e,e'pn)$ could considerably improve the comparison of theory and experiment.
\begin{figure}[ht]
\centerline{\includegraphics*[scale=.5]{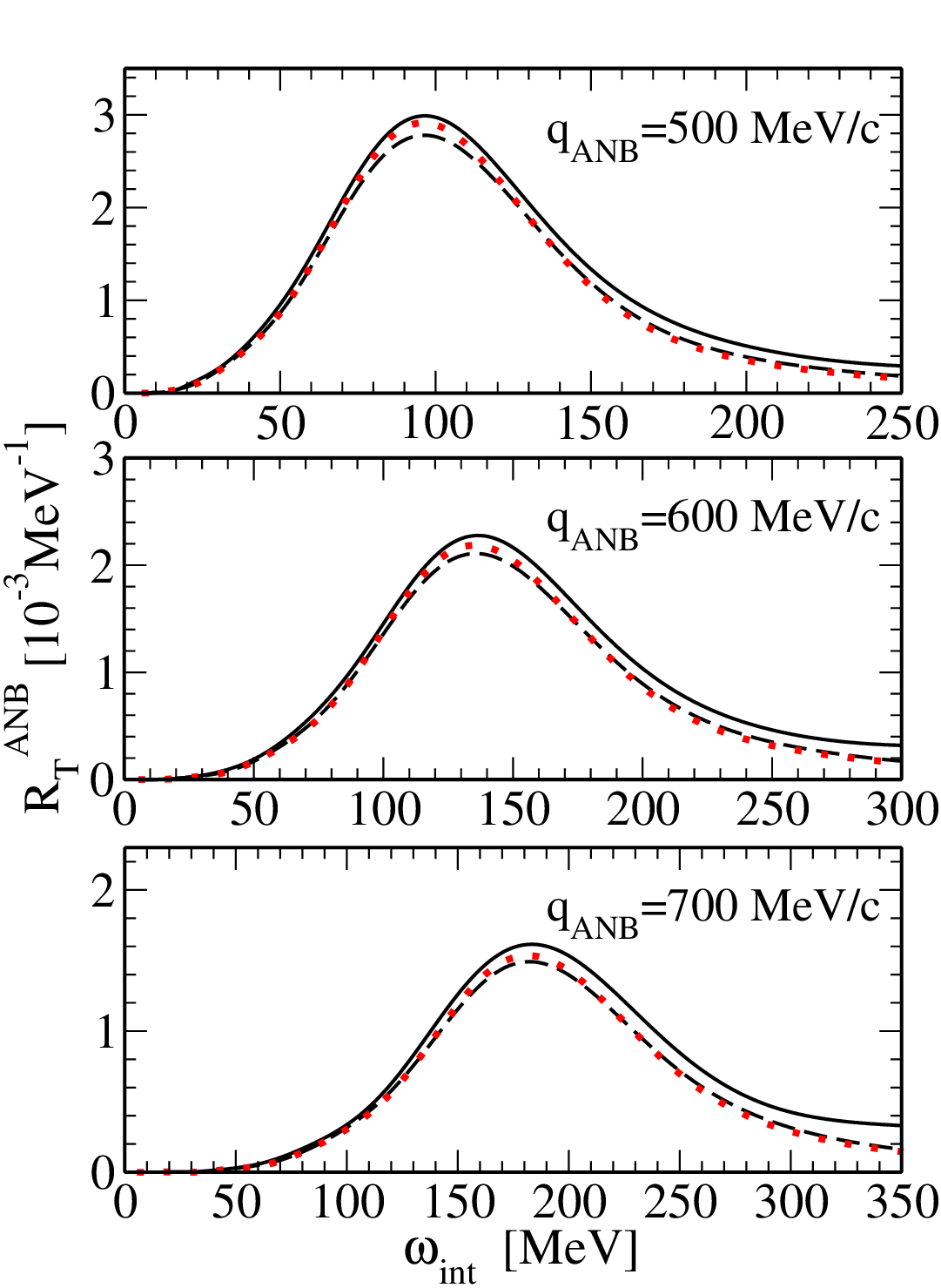}}
\caption{As in Fig.~3 but only transitions to final states with total isospin $T_f=3/2$ are taken into account.}
\label{T3half}
\end{figure}

We summarize our work as follows. We have calculated the transverse electron scattering response function at 500 MeV/c $\le q \le$ 700 MeV/c. For the nuclear current operator we have taken into account the non-relativistic one-body operator plus first-order relativistic corrections, meson exchange currents and currents involving the $\Delta$ isobar. This marks the first time our calculations
with quasi-elastic kinematics have included the $\Delta$ isobar. The calculation is made with the Lorentz integral transform method, which enables a rigorous inclusion of final state interactions. The transverse response function $R_T(q,\omega)$ is calculated in the ANB frame with a subsequent transformation to the lab frame. Relativistic corrections to the kinetic energy are considered by a two-fragment model,  introduced in our previous studies, which is particularly appropriate for the quasi-elastic peak region. The $\Delta$ current contribution enhances $R_T$ in the peak region by about 5\%. Though it is a rather moderate effect it improves the theoretical result leading to an excellent agreement with experimental data. In addition, and in agreement with the results of Ref.~\cite{Deltuva04} at lower $q$, we have shown that three-nucleon force effects and the $\Delta$ current contribution largely cancel each other in the peak region. Beyond the peak region $\Delta$ degrees of freedom become increasingly important, particularly for the isospin $T_f=3/2$ channel which contributes exclusively to three-body break-up reactions. This finding could explain why a recent study of the reaction $^3$He$(e,e'pn)$  (where the theoretical portion did not include $\Delta$ degrees of freedom) produced large differences between the theoretical and experimental results in the dip region.

\vskip 0.5cm
Acknowledgments of financial support are given to AuroraScience (L.Y.), to
the RFBR, Grant No. 10-02-00718 and RMES, Grant NO. NS-7235.2010.2 (V.D.E.),
and to the National Science and Engineering Research Council of Canada (E.L.T.).

\end{document}